\documentclass[11pt]{article}
\usepackage{moriond,epsfig,color}

\font\douzesf=cmss10 at 12pt

\def\be{{\douzesf  $^{\douzesf  7}$Be}}

\def\Al{$^{26}$Al}

\def\Ni{$^{56}$Ni}

\def\Na{$^{22}$Na}
\def\Co{$^{56}$Co}
\def\Ci{$^{57}$Co}
\def\Ch{$^{60}$Co}
\def\Fe{$^{56}$Fe}

\def\Fh{$^{60}$Fe}

\def\Ti{$^{44}$Ti}

\def\ra{$\rightarrow$}

\def\ga{$\gamma$}

\def\cs{cm$^{-2}$ s$^{-1}$}
\def\ms{M$_{\odot}$}
\def\zs{Z$_{\odot}$}

\def\Journal#1#2#3#4{{#1} {\bf #2}, #3 (#4)}

\def\PR{{\em Phys. Rev.}}

\def\APJ{{\em ApJ}}
\def\APJS{{\em ApJ Suppl.}}
\def\AA{{\em A\&A}}
\def\ARAA{{\em ARA\&A} }
\def\ANYAS{{\em Ann. NY Acad. Sci.}}
\def\MPS{{\em Meteoritics and Planetary Sci.} }
\def\PHR{{\em Phys. Rep.} }
\def\RMP{{\em Rev. Modern Phys.} }
\def\NAR{{\em New Astr. Rev.} }
\def\CJP{{\em Canadian Journal of Physics} }
\def\PASP{{\em Publ. Astr. Soc. of the Pacific} }
\def\Nat{{\em Nature}}
\def\CNPP{{\em Com. Nucl. Part. Phys.} }
\def\MNRAS{{\em Monthly Not. Royal Astr. Soc.} }

\def\ra{\rightarrow}

\def\be{\begin{equation}}
\def\ee{\end{equation}}
\def\bea{\begin{eqnarray}}
\def\eea{\end{eqnarray}}

\begin{document}
\vspace*{4cm}
\title{STELLAR RADIOACTIVITIES AND DIFFUSE GAMMA-RAY LINE EMISSION IN THE MILKY WAY}

\author{ N. PRANTZOS }

\address{Institut d'Astrophysique de Paris, 98bis Bd Arago,\\
75014 Paris, France}

\maketitle\abstracts{
After a short historical introduction to the field of \ga-ray line astronomy with
radioactivities, I present an overview of recent results concerning the massive star
yields of those radioactivities. I comment on the implications of  those
results (concerning long-lived radioactivities, like \Al \ and \Fh) for \ga-ray
line astronomy, in the light of past ({\it COMPTEL} \ and GRIS) and forthcoming 
({\it INTEGRAL}) observations. }

\section{Introduction}

Shortly after the discovery of the phenomenon of radioactivity, radionuclides 
revealed to be unique '''probes'' in our study of the cosmos and important
agents in its evolution (radioactive dating of the Earth, meteorites and
stars; radioactive heating of planetary and supernova interiors; radioactive 
origin
of abundant stable nuclei, like \Fe, and of
isotopic anomalies in meteorites, etc).

As most other stable nuclei, radionuclides are produced in stellar interiors
and ejected in the interstellar medium through stellar winds and explosions
(nova or supernova). In a few cases, concerning extra-solar objects,
the characteristic \ga-ray line signature 
of their radioactive decay has been detected and used as
a probe of a large variety of astrophysical sites; indeed, 
\ga-ray line astronomy with cosmic radioactivities has grown to a mature
astrophysical discipline in the last decade. See. e.g. Diehl and Timmes
1998, Arnould and Prantzos 1999, Kn\"odlseder and Vedrenne 2001, for
recent reviews; also, the proceedings of the {\it Astronomy with Radioactivities} 
Conference, organised every two years, nicely reflects the status of that
discipline (web site: {\tt  
http://www.mpe.mpg.de/gamma/science/lines/workshops/radioactivity.htm } ).

In this review I shall focus on radioactivities produced by massive stars
(SNII and WR stars); radioactivities produced by exploding white dwarfs 
(novae and SNIa) are  reviewed by Hernanz (this volume).

\section{A short history of stellar radioactivities and $\gamma$-ray line
astronomy}

The main theoretical ideas underlying $\gamma$-ray line astronomy emerged 
slowly in the 60ies, while observational evidence came 
only about 20 years later. This history is largely dominated by two rather 
independent ``programmes'' of research: an astronomical one, seeking for the 
explanation of the late lightcurves of supernovae, and a nucleosynthetic one, 
seeking for the origin of the most abundant heavy nucleus, \Fe. An 
exceptionally clear and vivid account of that history is given in the text of
Clayton (1999), on which much of this section is based.

In the early 50ies, the exponential decline of the late lightcurves of SNIa was
attributed to the radioactive decay of $^7$Be (Borst 1950) or $^{59}$Fe (Anders 1959)
or $^{254}$Cf 
(Anders 1959, Burbidge et al. 1956), all those nuclei having half-lives of 
$\sim$45-55 days. 
In his PhD thesis (1962), the mineralogist T. Pankey Jr 
suggested that \Fe \ is produced as unstable \Ni, and that the radioactive
chain \Ni$\ra$\Co$\ra$\Fe \ can explain the lightcurves of supernovae;
however, his suggestion went completely unnoticed by astronomers and nuclear
physicists alike. Indeed, up to the mid-sixties it was 
thought that \Fe \ is produced as such in stellar interiors (Hoyle 1946; 
Burbidge et al. 1957; Fowler and Hoyle 1964), through the so-called 
{\it ``e-process''}, despite the fact that the issue of its {\it ejection} in 
the interstellar medium (which might modify its abundance) 
was far from being clear.
The role of {\it explosive Si-burning}, leading to the production (and natural 
ejection from supernovae) 
of doubly-magic \Ni \ was clarified through semi-analytical calculations
of Bodansky et al. (1968), 
after hints from pioneering numerical nucleosynthesis calculations of Truran 
et al. (1966).
Based on those results, Colgate and McKee (1969) convincingly argued that
the radioactive chain \Ni$\ra$\Co$\ra$\Fe \ powers
the lightcurves of supernovae; as time goes on, an increasing percentage of 
that power escapes the SN ejecta (which become progressively more transparent
to $\gamma$-rays) and as a result the optical light curve declines more rapidly
(by a factor of 2 every $\sim$55 days) than the amount of \Co \ (half-life: 77
days).

The implications of those ideas for $\gamma$-ray line astronomy were studied 
in the 60ies at the Rice University, where Clayton and Craddock (1965) first 
calculated the expected $\gamma$-ray flux and spectrum from the Crab remnant,
on the assumption that the $^{254}$Cf hypothesis was correct; finding that
extremely large overabundances of other heavy elements (Os, Ir, Pt) should
be obtained in that case, they expressed doubts on the correctness of that
hypothesis. After this ``false-start'', the implications of \Ni \
production in Si-burning were fully clarified in the landmark paper of
Clayton,  Colgate and Fishman (1969), which opened exciting perspectives
to the field by suggesting that any supernova 
within the local group of  galaxies should be detectable in $\gamma$-ray lines.

In the 70ies D. Clayton identified most of the radionuclides of 
astrophysical interest (i.e. giving a detectable $\gamma$-ray line signal);
for that purpose he evaluated their average SN yields, by assuming that the
corresponding daughter stable nuclei are produced in their solar system 
abundances. Amazingly enough (or naturally enough, depending on one's point
of view) his predictions of average SNII radionuclide yields (Table 2 in
Clayton 1982) are in excellent agreement with modern yield calculations, based
on full stellar models and detailed nuclear physics (see Fig. 1). Only the
importance of \Al \ escaped Clayton's (1982) attention, perhaps because its
daughter nucleus $^{26}$Mg is produced in its stable form, making the
evaluation of the parent's yield quite uncertain. That uncertainty did not
prevent Arnett (1977) and Ramaty and Lingenfelter (1977) 
from arguing (on the basis of
Arnett's (1969) explosive nucleosynthesis calculations) that, even if only
10$^{-3}$ of solar $^{26}$Mg is produced as \Al, the resulting Galactic
flux from tens of thousands of supernovae (during the $\sim$1 Myr lifetime
of \Al) would be of the order of 10$^{-4}$ \cs.

In the case of \Al \  nature appeared quite generous, providing a \ga-ray
flux even larger than the optimistic estimates of Ramaty and Lingenfelter 
(1977): the HEAO-3 satellite detected the corresponding 1.8 MeV line from 
the Galactic  center direction at a level
of 4 10$^{-4}$ \cs (Mahoney et al. 1984). That detection, the first ever of
a cosmic radioactivity, showed that nucleosynthesis is still active in the 
Milky Way; however, the implied large amount of galactic \Al \ ($\sim$3 \ms \
per Myr, assuming steady state) was difficult to accomodate in conventional
models of galactic chemical evolution if SNII were the main \Al \ source
(Clayton 1984), since $^{27}$Al would be overproduced in that case; however, 
if the  ``closed box model'' assumption is dropped and {\it infall}
is assumed in the chemical evolution model, that difficulty
is removed, as subsequently shown by Clayton and Leising (1987).

Another welcome mini-surprise came a few years later, when the \Co \
\ga-ray lines were detected in the supernova SN1987A, a $\sim$20 \ms \
star that exploded in the Large Magellanic Cloud. On theoretical  grounds,
it was expected that a SNIa (exploding white dwarf of $\sim$1.4 \ms \ that
produces $\sim$0.7 \ms \ of \Ni) would be the first to be detected in \ga-ray 
lines;
indeed, the large envelope mass of SNII ($\sim$10 \ms) allows only small 
amounts
of \ga-rays to leak out, making  the detectability of such objects problematic
(Woosley et al. 1981, Gehrels et al. 1987).
Despite the intrinsically weak \ga-ray line emissivity of SN1987A, the 
proximity
of LMC allowed the first detection of the tell-tale \ga-ray line signature from
the famous radioactive chain  \Ni$\ra$\Co$\ra$\Fe, thus confirming a 25-year 
old
conjecture (namely, that the abundant \Fe \ is produced in the
form of radioactive \Ni).

Those discoveries laid the observational foundations of the field of \ga-ray
line astronomy with radioactivities. The next steps were made in the 90ies, 
thanks
to the performances of the Compton Gamma-Ray Observatory (CGRO). First, the
{\it OSSE} instrument aboard CGRO detected the \Ci \ \ga-ray lines from SN1987A
(Kurfess et al. 1992); the determination of the
abundance ratio of the isotopes with mass numbers
56 and 57 offered a unique probe of the physical conditions in the innermost
layers of the supernova, where those isotopes are synthesized 
(Clayton et al. 1992). On the other hand, the {\it COMPTEL} \ instrument 
mapped the
Miky Way in the light of the 1.8 MeV line and found irregular emission
along the plane of the Milky Way and prominent ``hot-spots'' in directions
tangent to the spiral arms (Diehl et al. 1995); that map implies
that massive stars (SNII and/or WR) are at the origin of galactic \Al \ (as
suggested by Prantzos 1991, 1993) and not an old stellar population like
novae or AGB stars. Furthermore, {\it COMPTEL} \  detected the 1.16 MeV line of
 radioactive
\Ti \ in the Cas-A supernova remnant (Iyudin et al 1994); that discovery
 offered
another valuable estimate of the yield of a radioactive isotope produced
in a massive star explosion (although, in that case the progenitor star mass
is not known, contrary to the case of SN1987A).

After that short historical introduction to the field of \ga-ray line 
astronomy,
we turn in the next section into a discussion of the theoretically predicted
yields of radioactivities from massive stars, the associated uncertainties and
the relevant observational constraints.

\section{Stellar Radioactivities: Yields, constraints, detectability}

\subsection{Overview}

All nuclei (except for the primordial isotopes of H and He and those  
of Li, Be and B) are thermonuclearly
synthesized in the hot and dense stellar interiors, which are opaque to
\ga-rays. Released \ga-ray photons interact with the surrounding material 
and are 
Compton-scattered down to X-ray energies, until they are photoelectrically
absorbed and their energy is emitted at longer wavelengths. To become 
detectable,
radioactive nuclei have to be brought to the surface (through vigorous
convection) and/or ejected in the interstellar medium, either through stellar
winds (AGB and WR stars) or an explosion (novae or supernovae). Their 
detection 
provides then unique information on their production sites.

\begin{table*}[t]
\caption{Important stellar radioactivities for gamma-ray line astronomy
\label{fig:Table1}}
\vskip -2.8 cm
\psfig{figure=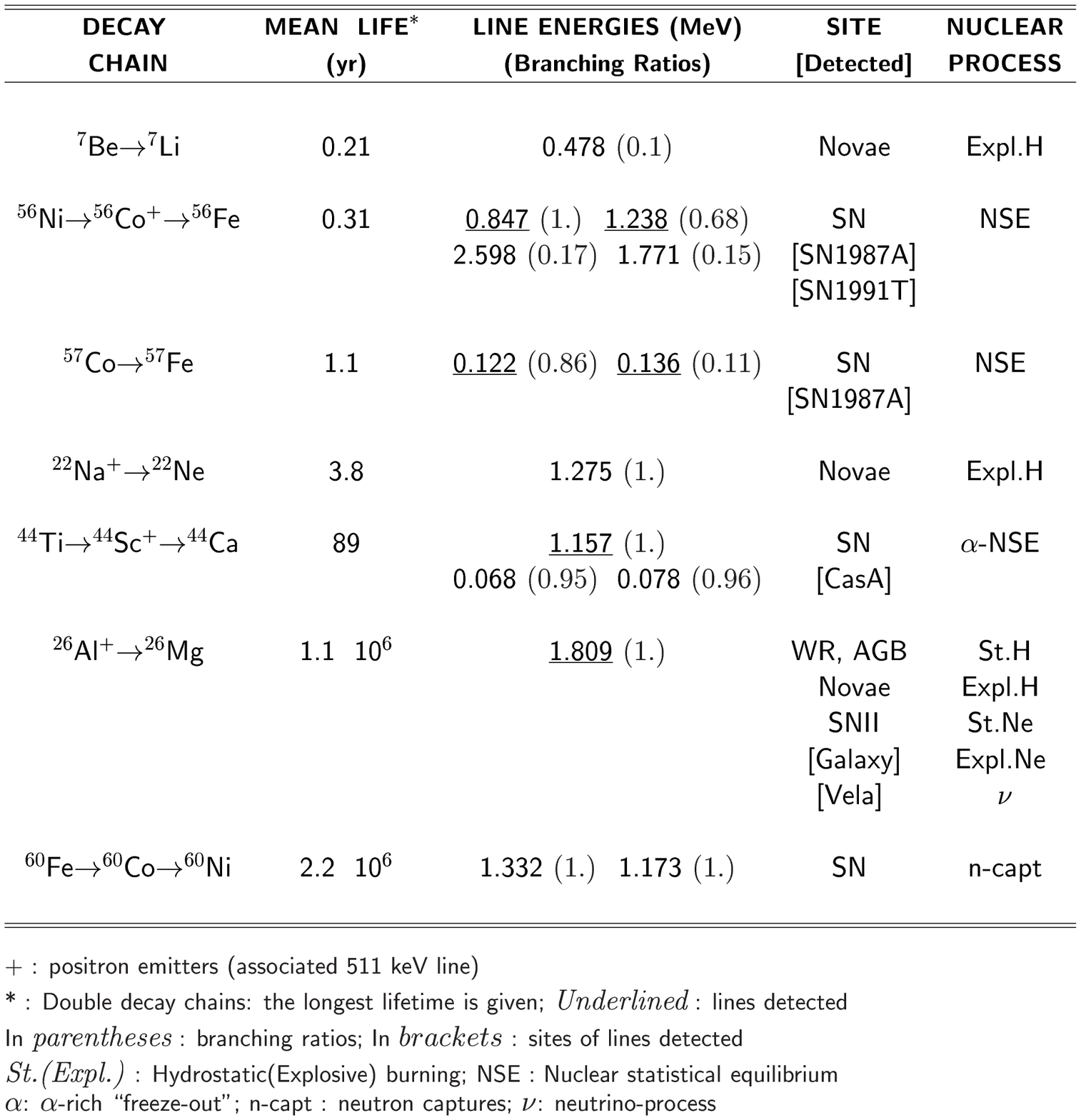,width=\textwidth,height=19.cm}
\vskip -4.4 cm
\end{table*}

The intensity of the escaping \ga-ray lines gives important
information on the yields of the corresponding isotopes and the
physical conditions (temperature, density, neutron excess etc.) in the
stellar zones of their production, as well as on other features of the
production sites (extent of convection, mass loss, hydrodynamic instabilities,
position of the ``mass-cut'' in SNII, etc.).
The shape of the \ga-ray lines reflects
the velocity distribution of the ejecta, modified by the opacity along
the line of sight and can give information on the structure
of the ejecta (see e.g. Burrows 1991 for the potential of \ga-ray lines
as a tool of supernova diagnostics). Up to now, only the 0.847 MeV  \Co \ line
from SN1987A and the 1.8 MeV \Al \ line 
from the inner Galaxy have been resolved (both with the same
instrument, the balloon borne GRIS spectrometer), but their ``message'' 
is not quite understood yet.

Obviously, radionuclides of interest for \ga-ray line astronomy are those
with high enough yields and short enough lifetimes for the emerging \ga-ray lines
to be detectable. On the basis of those criteria, Table 1 gives the most important
radionuclides (or radioactive chains) for \ga-ray line astronomy, along with
the corresponding lifetimes, line energies and branching ratios, production sites
and nucleosynthetic processes.\footnote{The 511 keV line of e$^+$-e$^-$
annihilation is, in fact, the first
\ga-ray line ever detected (Johnson et al. 1972), although its origin (probably related
to the radionuclides of Table 1) and spatial distribution in the Galaxy 
are not well understood yet (see
Kinzer et al. 2001, and references therein).}

When the lifetime of a  radioactive nucleus  is not very large w.r.t.
the timescale between two nucleosynthetic events in the Galaxy,
those events are expected to be seen as point-sources in the light
of that radioactivity. In the opposite case a diffuse emission along
the Galaxy is expected from the cumulated emission of hundreds
or thousands of sources.  Characteristic timescales between two
explosions  are $\sim$1-2 weeks for novae 
(from their estimated Galactic frequency of
$\sim$30 yr$^{-1}$, Della Vale and Livio 1995),
$\sim$50-100 yr for SNII+SNIb and $\sim$200-400 yr
for SNIa (from the corresponding Galactic frequencies of $\sim$2
SNII+SNIb century$^{-1}$ and $\sim$0.25-0.5 SNIa century$^{-1}$,
Tammann et al. 1994, Cappellaro et al. 1997).
Comparing those timescales to the decay lifetimes of Table 1
one sees that in the case of the long-lived \Al \ and \Fh \
a diffuse emission is expected; the spatial profile of that emission
should reflect the Galactic distribution of the
underlying sources.  All the other radioactivities of Table 1
should be seen as point sources in the Galaxy except, perhaps,
$^{22}$Na from Galactic novae; indeed, the most prolific \Na \
producers, O-Ne-Mg rich novae, have a
frequency $\sim$1/3 of the total (i.e. $\sim$10 yr$^{-1}$),
resulting in $\sim$40 sources active  in the Galaxy during
the 3.8 yr lifetime of $^{22}$Na.

\begin{figure}[t]
\begin{center}
\psfig{figure=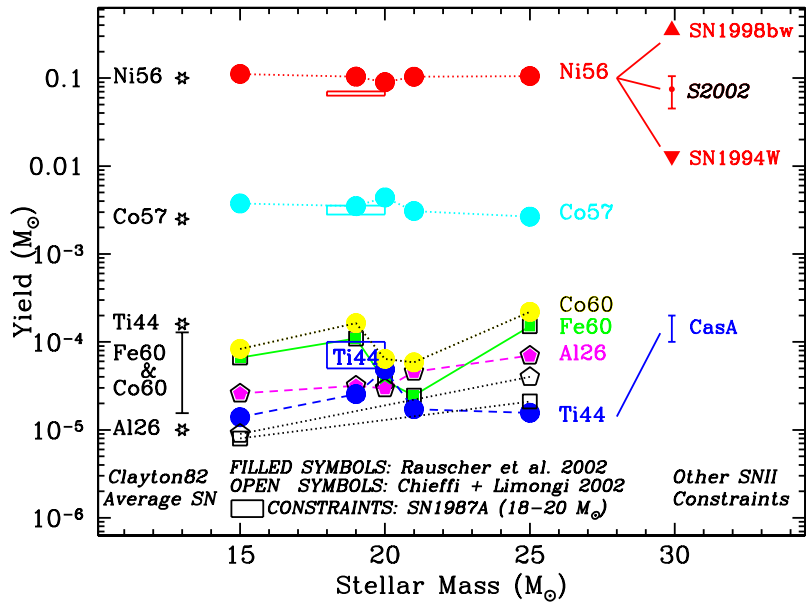,angle=-90,width=0.7\textwidth}
\caption{Yields of radioactive nuclei from massive exploding stars. 
{\it On the left},  Clayton's (1982) predictions for average SN yields
are given (based on 
the assumption that the corresponding stable daughter isotopes are produced 
in their solar abundances as radioactive progenitors).
{\it In the middle}, recent results are plotted as a function of stellar mass:
{\it filled symbols} are from Rauscher et al. (2002) and {\it open symbols} from Chieffi
and Limongi (2002), while {\it open parallelograms} indicate observational constraints
from SN1987A (see Sec. 3.3). {\it On the right}, other observational constraints are
given (the stellar mass in the abcissa is irrelevant in that case):
 those for \Ni \ are for extragalactic SNII from Sollerman (2002, S2002), 
with upper and lower limits for SN1994W and SN1998bw, respectively;
the CasA results concern estimates of ejected \Ti, detected through its 0.068, 0.078 and
1.16 MeV lines (see Sec. 3.3)
\label{fig:Fig1}}
\end{center}
\end{figure}

\subsection{Yields}

Yields of radioactive isotopes produced in SNII
are displayed in Fig. 1. On the left part of the
diagram, Clayton's (1982) ``educated guess'' of those yields is
presented for illustration purposes; as discussed in Sec. 2, it
is in excellent agreement with modern yield calculations. 

In Fig. 1 it appears
that the stellar mass does not affect substantially those yields; at least
in the 15-25 \ms \ mass range, yields do not vary by more than a factor
of $\sim$2-3 (notice, however, that they do not always behave monotonically
with mass). Unfortunately, the uncertainties in those yields are difficult to
quantify at present, because of the many factors involved: nuclear physics 
(for instance, the $^{12}$C($\alpha,\gamma$) rate or n-capture and n-production cross sections), 
convection and mass loss prescriptions, position  of the mass-cut, neutrino spectra (for some
nuclei that may receive conribution from neutrino-induced nucleosynthesis) etc.
Taking all those uncertainties into account, it is safe to assume that 
theoretical yields at present are uncertain by at least a factor of 2 (and, quite
probably, by much larger factors). In particular, the yield of all Fe-peak 
radioactivities (including \Ti) are quite sensitive to the position of the mass-cut;
some discussion on relevant constraints is given in Sec. 3.3. Here we proceed to a
comparison between results of 2 recent calculations, by Rauscher et al. (2002 or 
RHHW2002) and Chieffi and Limongi
(2002 or CL2002), performed with state-of-the-art stellar evolution models
(including mass loss and a simulation of the explosion) and extended nuclear
reaction networks with updated physics. These results illustrate well 
current uncertainties for \Al \ and \Fh,
two radioactivities produced outside the stellar Fe-core.

- In the case of \Al, the overall agreement is rather good: the  RHHW2002 yields
are larger by a factor of 2.5 on average than those of CL2002, the difference been 
more pronounced in the 15 \ms \ star than in the 25 \ms \ case. The two calculations
converge in the more massive stars, where \Al \ production is dominated by pre-explosive
nucleosynthesis in the Ne and H shells. In lower mass stars \Al \ production is
dominated by explosive Ne-burning; several factors may 
then explain the differences between the two calculations:
the detailed pre-supernova structure
through which the shock-wave runs; the amount of seed nuclei ($^{23}$Na, $^{25}$Mg etc)
which are products of C-burning and depend thereoff on the carbon abundance left off from
He-burning, that is on the $^{12}$C($\alpha,\gamma$) reaction rate;  
the $\nu$-induced nucleosynthesis (included in RHHW2002 but not in CL2002), etc.

- In the case of \Fh \ the situation is not as satisfactory as for \Al. \Fh \ is 
mainly produced by explosive Ne-burning, through neutron captures on stable \Fe \ and
$^{58}$Fe; its yield depends on the available amount of $^{22}$Ne, which releases those
neutrons through $^{22}$Ne($\alpha$,n), as well as on available $^{58}$Fe. There is a 
factor of $\sim$10 difference between the two calculations, for both the 15 \ms \ and
the 25 \ms \ stars. An explanation of such a large difference appears difficult,
especially when the non-monotonic  behaviour of the  RHHW2002 yields of \Fh \ with
stellar mass is taken into account: according to  RHHW2002, the $\sim$20 \ms \ region
marks the transition from exoergic convective carbon burning (for M$<$20 \ms)
to stars where energy production from central C-burning just compensates for
neutrino losses (M$>$20 \ms); the effect of that transition on the \Fh \ yields
has not been investigated yet. Notice that the  \Fh \ yields of  RHHW2002 are much
larger than those of the previous calculations of that same group 
(Woosley and Weaver 1995). Notice also that  the  \Fh \ yields of  RHHW2002 are larger
than the corresponding ones of \Al, a situation that is not encountered either
in CL2002 or in Woosley and Weaver (1995).

\subsection{Constraints}

The issue of the \Fh \ and \Al \ yields in massive stars is of importance, in view
of current observational constraints and forthcoming {\it INTEGRAL} measurements (see 
Sec. 4.2). Fig. 1 displays some other observational constraints on SNII
radioactivities, obtained for SN1987A (paralellograms for \Ni, 
\Ci \ and \Ti, for a 18-20 \ms \ star) and for other supernovae (on the right of the
figure; the corresponding stellar mass is irrelevant in the latter case).

In the case of SN1987A, the \Ni \  yield (0.07 \ms) is obtained through extrapolation 
of the supernova lightcurve, assumed to be powered by \Co \ decay, to the day
of the explosion (e.g. Arnett et al. 1989). The yield of \Ci \ is obtained
in three different ways: a) through the measured intensity of the 0.122 Mev line 
of \Ci \ and
assuming a low optical depth for those photons; b) through the study of the
late bolometric lightcurve of SN1987A and assuming that it is dominated by \Ci \ 
decay at days 1100-2000 (this analysis is far less straightforward than in the case
of \Ni); c) through an analysis of the infrared emission lines of the ejecta. All
those methods converge to a value of \Ci \ mass of $\sim$ 3 10$^{-3}$ \ms \
(see Fransson and Kozma 2002). Finally, the yield of \Ti \ is evaluated through 
methods (b) and (c), albeit with substantial difficulties, due to the complex
physics of supernova heating and coooling involved and the role of positrons; 
current estimates give values in the 0.5-2 10$^{-4}$ \ms \ range (Fransson and Kozma 2002),
while Sollerman (2002) suggests an upper limit of 1.1 10$^{-4}$ \ms.

These observational constraints compare rather well with theoretical predictions
for 18-20 \ms \ stars (the estimated progenitor mass of SN1987A, on the basis
of its optical luminosity, e.g. Arnett et al. (1989)). Notice, however, that
model results in Fig. 1 correspond to stars calculated with initial metallicity Z=\zs, while
the progenitor of SN1987A presumably had  LMC metallicity, namely Z$\sim$0.3 \zs.
Notice also that Thielemann et al. (1996) obtain  a larger \Ti \ yield for the
20 \ms \ star (1.7 10$^{-4}$ \ms), due to a difference in the way of simulating
the explosion: the ``thermal bomb'' they use leads to a larger entropy and more
important $\alpha$-rich freeze-out than in the case of the piston-driven explosion
adopted by RHHW2002. Such a high \Ti \ yield is marginally detectable by {\it INTEGRAL}
(see next section).

Data on the right of  Fig. 1 concern \Ni \ yield estimates for extragalactic SNII.
Based  on a sample of 8 SNIIP (the ``standard'' SNII, with a ``plateau'' 
in the optical lightcurve) and assuming a bolometric correction similar 
to the one of SN1987A, Sollerman (2002) finds a mean value of 0.075 \ms \ with 
a standard deviation of 0.03 \ms. He notices, however, that  SNII with 
much lower and higher yields than the ``canonical'' one have also been found. 
In the former case belong SN1994W: the extremely
rapid fading of its lightcurve suggests a \Ni \ yield lower than 0.015 \ms. 
On the other hand, SN1998bw is the most \Ni-rich supernova today: detailed 
modelling of its late emission requires \Ni \ yields of 0.5-0.9 \ms, and simple 
arguments lead to a lower limit of 0.3 \ms \ (Sollerman et al. 2002).
Thus, it appears that the \Fe \ yield of massive stars is far from being 
a ``universal constant'' of $\sim$0.075
\ms, a fact that may have interesting implications for stellar models as well as
galactic chemical evolution,
especially concerning the observed scatter of abundance ratios in halo stars
(Ishimaru et al. 2002).

Finally, the \Ti \ yield of CasA  is
inferred from the 1.16 MeV line flux of $^{44}$Sc decay detected by {\it COMPTEL} \
(3.3$\pm$0.6 10$^{-5}$ \cs) and the CasA distance (3.4 kpc) and age (320 yr) and amounts
to $\sim$1.7 10$^{-4}$ \ms (Iyudin et al. 1999).
An independent evaluation of the \Ti \ yield in CasA came
recently, through detection of the low energy decay lines of \Ti \ by 
{\it Beppo-SAX}: the detected flux at 68 and 78 keV implies a \Ti \ mass of 
1-2 10$^{-4}$ \ms, depending on the modelisation of the underlying continuum
spectrum (Vink et al. 2001; Vink and Laming this volume). 
These yields are larger than the average \Ti \ yields of RHHW2002 (see Fig. 1), typically
by a factor of $\sim$3, but compatible with those of Thielemann et al. (1996).
Notice, however, that these estimates suffer from
uncertainties related to the ionisation state of the SN remnant; an ionised medium
could slow down the electron-capture decay of that radionuclide and explain the observed flux with
a smaller yield (see Mochizuki et al. 1999).

\subsection{Detectability}

For tutorial purposes, we 
present in Fig. 2 a schematic view of the \ga-ray line emissivity of a ``typical''
SNII, over three different timescales: 10 years, 10 centuries and a few Myrs.
The figure is based on the yields of Fig. 1 and is calculated by
assuming a SN1987A-like opacity for the ejecta.

\begin{figure}[t]
\begin{center}
\psfig{figure=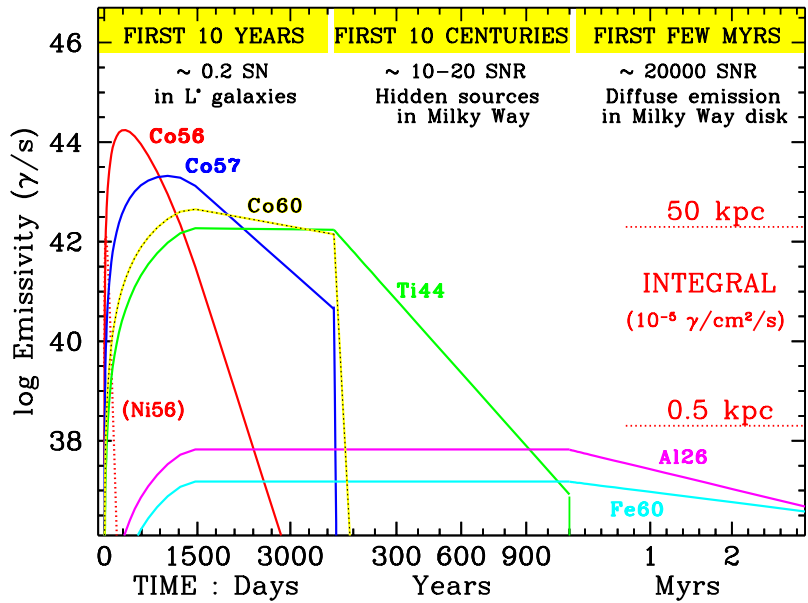,angle=-90,width=0.7\textwidth}
\caption{Schematic evolution of the gamma-ray line emissivity of a 
``typical'' SNII, over three
different timescales (10 years, 10 centuries, few Myrs).
The figure is based on the yields of Fig. 1 and on an opacity for the SN ejecta
similar to the  one in the case of SN1987A. On the right are shown maximum detection
distances for {\it INTEGRAL}, corresponding to those emissivities; {\it INTEGRAL} may detect Ti44
up to 50 kpc (e.g. from SN1987A) or Al26 from a nearby event, closer than 0.3 kpc  
(e.g. from objects in the Vela region, see text).
\label{fig: Fig2}}
\end{center}
\end{figure}

Notice that, if  the RHHW2002 yields of $^{60}$Co are correct, the \Ch \
lines might dominate
the \ga-ray line emission of the SN for a couple of years, between 5 and 8 
years after the explosion; that  possibility was suggested by Clayton (1982) for very
young SN remnants in the Milky Way. Unfortunately, the expected flux from SN1987A
was below the sensitivity limits of instruments aboard CGRO and it will
also be below the detection threshold of {\it INTEGRAL} (which is launched $\sim$15 years
after the explosion, while \Ch \ has a mean life of 7.6 yr). The role of \Ch \ 
for the
late lightcurve of SN1987A was studied in Timmes et al. (1996). It may well be that
the current difficulties in modelling  the late bolometric lightcurve of that supernova
 and its infrared line emissivity (see previous section) may be, at least partially, due
to an inadequate account of  the energy input from that isotope.

The expected 1.16 MeV \ga-ray line flux from \Ti \ in SN1987A
($\sim$10$^{-5}$ \cs) lies at the detection limit of
{\it INTEGRAL} and will be one of the prime targets of the SPI instrument aboard that
satellite. Even a 3-$\sigma$ upper limit would bring important information on the
position of the mass-cut and the explosion mechanism of that supernova, since 
\Ti \ yield is more sensitive to the mass-cut  than other isotopes (e.g.
Timmes et al. 1996). On the other
hand, Fig. 2 reveals also that \Ti \ from centuries-old SN remnants in the Milky Way
should be detectable by {\it INTEGRAL}; here again, a positive detection will reveal
hitherto unknown Galactic SN remnants, while a negative result is expected to 
place interesting constraints on the frequency of the production sites of that isotope
and on the corresponding yields. Indeed, on the basis of Woosley and Weaver (1995)
yields Timmes et al. (1996) estimate that, in order to explain the solar abundance of 
$^{44}$Ca, one has to invoke either a higher SN frequency in the Galaxy or high \Ti
\ yields or production of $^{44}$Ca in rare events, like sub-Chandrasekhar mass SNIa. 
An analysis of {\it COMPTEL} \ map of the inner Galaxy in the light of 1.16 MeV suggests that
the first two possibilities should be excluded, otherwise more and/or brighter 
``hot-spots'' than actually observed should be found by {\it COMPTEL} \ (The et al. 2000). 
In that respect, it is interesting to notice that tantalizing hints for \Ti \ emission
from the nearby source GRO J0852-4642, a previously unknown supernova
remnant, were recently reported (Iyudin et al. 1998, Aschenbach et al. 1999; but, see also
Sch\"onfelder et al. 2000).

Long -lived radioactivities are difficult to detect from individual sources, 
even with next generation instruments. For instance, in the case of \Al, an exceptionally 
close site (closer than  $\sim$0.3 kpc) is
required for its 1.8 MeV line to be detectable by {\it INTEGRAL}; the Vela region
might offer just such a chance, in view of some intriguing hints from 
{\it COMPTEL} \  observations (see Sec. 4.4).
In the following we shall focus on the long-lived 
radioactivities \Al \ and \Fh. During
their $\sim$Myr lifetimes the collective emission from tens 
of thousands of sources
gives rise to a diffuse emission along the plane of the 
Milky Way; only the \Al \ emission has been detected up to now.

\section {Diffuse \ga-ray line emission from long-lived \Al \ and \Fh \ in the Milky Way}

\subsection{Overview}

{\it COMPTEL} \  is the only instrument with imaging capabilities that detected
the Galactic 1.8 MeV line emission (Fig. 3).
The data shows clearly a diffuse, irregular, emission  along the
Galactic plane, allowing to eliminate: i) a unique point source
in the Galactic centre and/or a nearby local bubble in that direction;
ii) an important contribution of the Galactic bulge, signature of an
old population and iii) any class of sources involving a large number
of sites with low individual yields (like nova or low mass AGB stars),
since a smooth flux distribution is expected in that case (Diehl et al. 1995). 
Identification of some of the observed features (``hot-spots'') with tangents to
spiral arms seems quite plausible and suggests that massive stars are at the origin
of \Al \ (Prantzos and Diehl 1996).

\begin{figure}[t]
\begin{center}
\psfig{figure=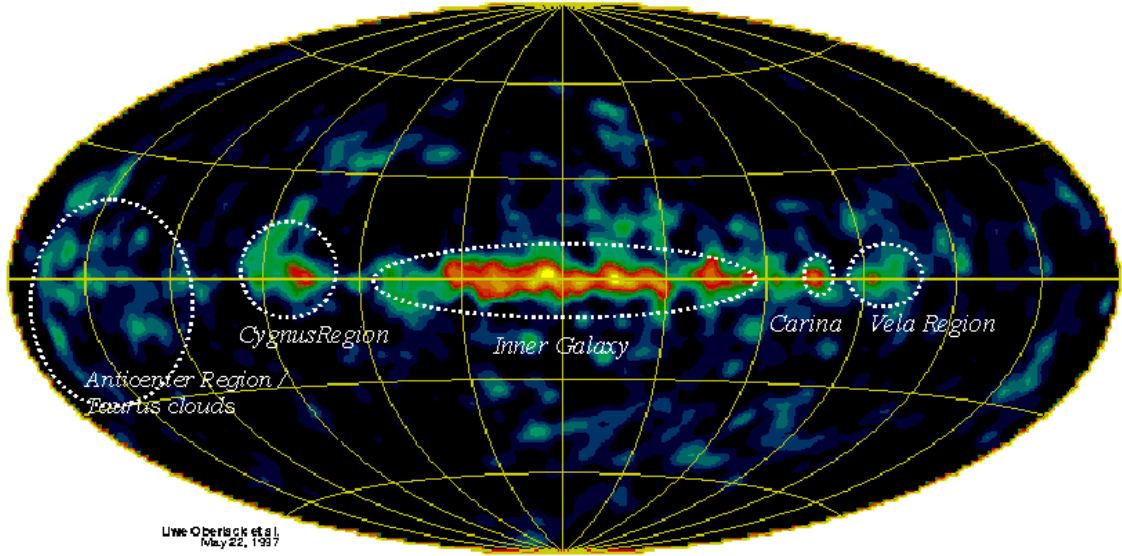,angle=-90,width=0.95\textwidth}
\caption{Sky map in the light of the 1.8 MeV line of Al26, according to the analysis
of the 5-year {\it COMPTEL} \  data by Oberlack (1997); several ``hot-spots'' in the
inner Galaxy and along the line of sight to spiral arms (Cygnus, Carina) are 
clearly identified.
\label{fig:Fig5}}
\end{center}
\end{figure}

Estimates of the galactic mass of \Al \ rely on assumptions about the spatial distribution
of the underlying sources. All plausible disk
models tested by the {\it COMPTEL} \  team yield a mass of $\sim$2
\ms. Introducing a spiral structure to the axisymmetric disk models
improves the fit to the data and implies that between 60 and 100 $\%$ of the \Al \ may lie
on the spiral arms (Diehl et al. 1998). It should be noticed that the derived spatial
distribution of \Al \ depends on the method of analysis. As shown by Kn\"odlseder et al. 
(1999) some imaging analysis methods  lead to all-sky maps with more pronounced 
localised features than some others; still, the irregular nature of the 1.8 MeV emission
along the Galactic plane and the localised ``hot-spots'' are revealed by all imaging
methods, in a statistically significant way (Pl\"uschke et al. 2001a).

\subsection{Sources of \Al \ and the role of \Fh}

The \Al \ yields presented in Fig. 1 concern massive stars exploding as SNII.
Even more massive stars ($>$30 \ms) may produce substantial amounts of
\Al \ during central ({\it hydrostatic})
H-burning and eject them through their powerful
stellar winds, in the WR stage (Prantzos and Cass\'e 1986);  the WR yields
are relatively well determined (e.g. Meynet et al. 1997), but the {\it explosive}
yields of those stars (which ultimately explode as SNIb) are very poorly known
at present.

\begin{figure}[t]
\begin{center}
\psfig{figure=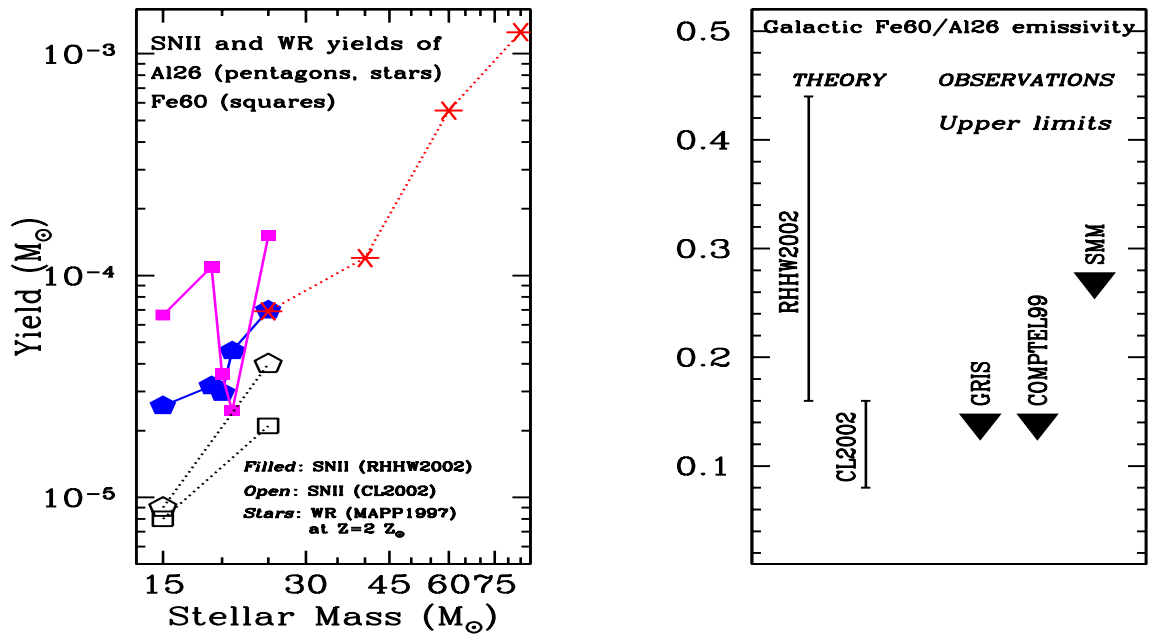,height=\textwidth,width=0.55\textwidth,angle=-90}
\caption{{\it Left:} Yields of Al26 ({\it pentagons} from SNII and {\it stars} from WR)
and Fe60 ({\it squares} from SNII) as a function of stellar mass, according to
Rauscher et al. (2002, RHHW2002, {\it filled symbols}),
Chieffi and Limongi (2002, CL2002, {\it open symbols}) and
Meynet et al. (1997, MAPP1997, {\it stars}); SNII yields are for Z$_{\odot}$ stars
(Al26 and Fe60 yields of SNII have a mild dependence on stellar metallicity), while
WR yields are for Z=2 Z$_{\odot}$ stars (the average metallicity of the inner Galaxy). 
{\it Right}: Flux ratio of the Fe60 to Al26 gamma-ray lines. The range of theoretical 
results corresponds to the yieds displayed in the left panel: the upper limit corresponds
to the average contribution of SNII alone, while for the lower limit the Al26 contribution
of WR stars is also taken into account; clearly, the yields of Rauscher et al. (2002)
violate the observed upper limits on the right of the figure, while those of Chieffi and Limongi
(2002) do not violate those limits, especially if  a substantial  WR contribution 
is assumed; notice, however, that the CL2002 yields produce only $\sim$0.4 \ms \ of \Al/Myr.
Observational upper limits are from GRIS (Naya et al. 1998), {\it COMPTEL} \ (Diehl 2000).
\label{fig:radish}}
\end{center}
\end{figure}

\begin{figure}[t]
\begin{center}
\psfig{figure=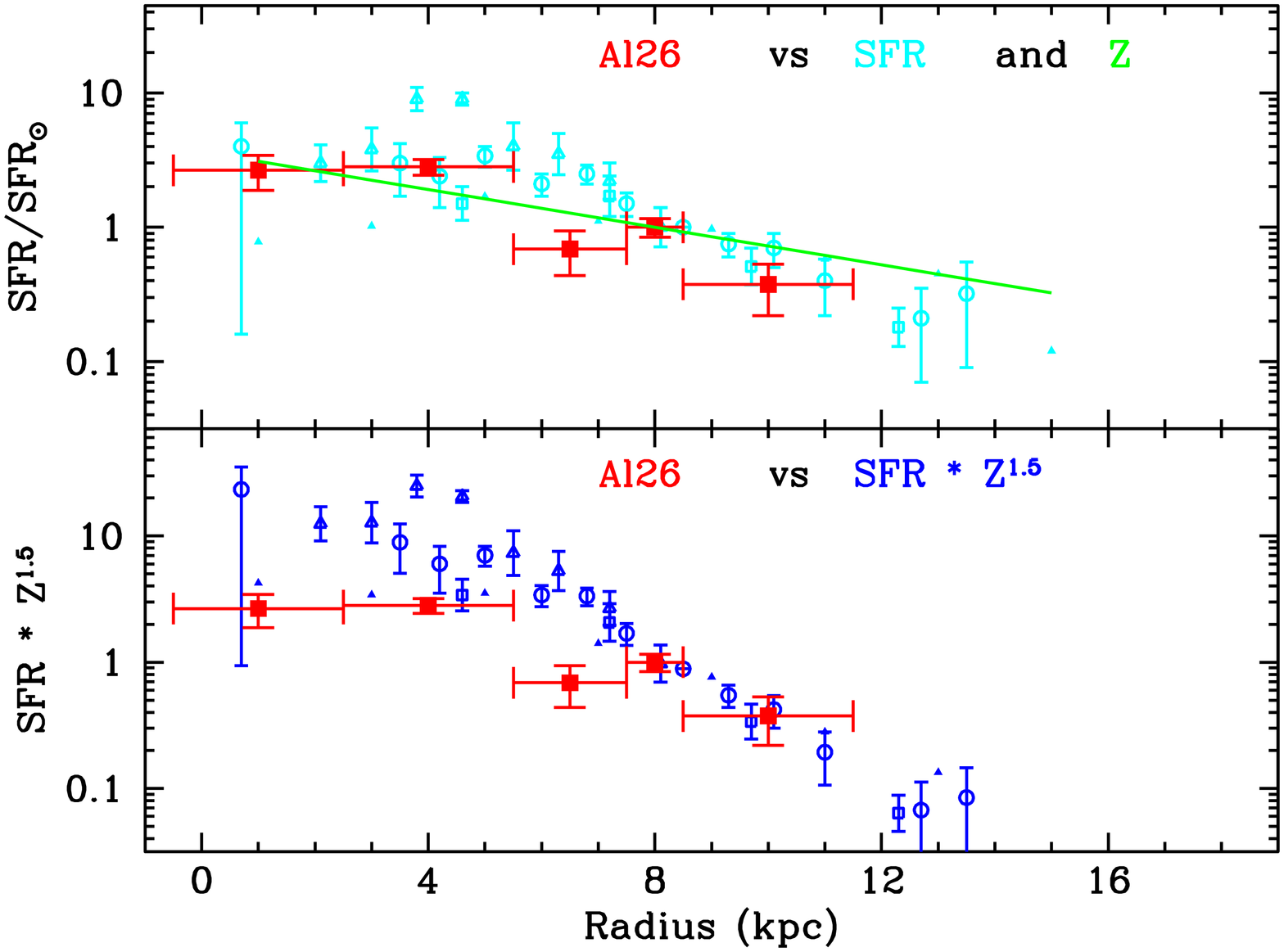,angle=-90,width=0.7\textwidth}
\caption{Radial distributions of Al26, star formation rate (SFR) and metallicity (Z)
in the Milky Way disk. {\it Upper panel:} Data points with vertical error bars correspond
to various tracers of the SFR, while the galactic metallicity profile of oxygen (with
a gradient of dlog(O/H)=-0.07 dex/kpc) is shown by a solid line;  the
Al26 profile, after an analysis of {\it COMPTEL} \ data by Kn\"odlseder (1997), is shown 
(in relative units)
by data points with vertical and horizontal error bars (the horizontal ones
correspond to the adopted radial binning). {\it Lower panel:} If galactic Al26 originates
mostly from WR stars, its radial distribution should scale with SFR * Z$^{1.5}$, since
the Al26 yields of WR stars scale with Z$^{1.5}$ (Meynet et al. 1997); however, the
observed Al26 distribution (same points as in upper panel) is flatter than the expected
one in that case. 
\label{fig:Fig5}}
\end{center}
\end{figure}

Under the most favorable conditions (highest possible yields for SNII allowed by
current uncertainties; accounting for the strong metallicity dependence of WR yields,
which favours sources in the inner Galaxy; adopting a mildly steep IMF, i.e. with
the Salpeter slope of -1.35 instead of the Scalo slope of -1.7), it turns out that
both SNII and WR can account for $\sim$2 \ms/Myr of \Al (e.g. Prantzos and Diehl 1996).
It may well be that both classes of
sources contribute equally to the Galactic \Al (a coincidence not ``stranger'' than
the quasi-equality between the solar abundances of s- and r- elements, or between
the contributions of the dark matter and dark energy to the density of the Universe).
However, it is interesting to see whether independent constraints can be used to
distinguish between the SNII and WR contributions and identify a dominant component
(assuming that there is one).

One such constraint is the flux ratio of the \ga-ray lines of \Fh \ (1.17 and 1.33 MeV) 
and \Al \ (1.8 MeV). Indeed, \Fh \ is predicted to be co-produced with \Al \  in
SNII (in almost the same zones and in similar amounts, Fig. 1), but not
in WR stars. If SNII dominate galactic \Al \ production, an  important
\Fh \ emission is then expected (flux ratio: ${^{60}Fe}\over{^{26}Al}$ = 
${Y_{60}/60/\tau_{60}}\over{Y_{26}/26/\tau_{26}}$, where Y represent yields averaged
over the IMF and $\tau$ the corresponding decay lifetimes); 
if WR stars are dominant, the \ga-ray
line flux ratio of \Fh/\Al \ is expected to be extremely low.

The  ${^{60}Fe}\over{^{26}Al}$ ratio of SNII depends on stellar models (and slightly
on the IMF). The Woosley and Weaver (1995) yields lead to a flux ratio of 
0.16 (Timmes et al. 1995), 
and so do the recent ones of CL2002 (note that the absolute 
\Fh \ and \Al \ yields of CL2002 are $\sim$ 4 times lower than those of WW1995); 
however, the most recent results of the Santa
Cruz group (RHHW2002) lead to a surprisingly large flux ratio 
${^{60}Fe}\over{^{26}Al}$$\sim$0.4. On the other hand,
current observational upper limits, obtained
by GRIS (Naya et al. 1998) and {\it COMPTEL} \ (Diehl 2000) are close to 0.15.
It appears then that: a) the RHHW2002 yields can produce $\sim$1 \ms \ of Galactic \Al,
but should be excluded by the non-detection of \Fh; b) the CL2002 yields produce
a Galactic \Al \ mass of $\sim$0.4 \ms, too low to explain the detected 1.8 MeV flux.
Taken at face value, the most recent SNII yields apparently exclude SNII as
dominant sources of Galactic \Al. Does this mean that WR stars constitute a viable
alternative?

WR stars can indeed provide $\sim$2 \ms/Myr of \Al \
(Meynet et al. 1997), provided that the strong metallicity dependence
of the yields is taken into account. Moreover, Kn\"odlseder (1999) showed that
a map of the ionising power from massive stars (derived from
the COBE data, after correction for synchrotron contribution) corresponds
to the 1.809 MeV map of galactic \Al \ in all significant detail; assuming a
standard stellar initial mass function, his calculation reproduces consistently
the current galactic supernova rate and massive star population from both maps,
and suggests that most of \Al \ is produced by WR stars of high metallicity in the
inner Galaxy. Finally, Kn\"odlseder et al. (2001) point out that one of the
prominent ``hot-spots'' in the {\it COMPTEL} \ 1.8 MeV sky-map, the Cygnus region, is an
association of massive stars with no sign of recent supernova activity.
All these observational and theoretical indices favour WR stars as dominant
\Al \ contributors. However, in that case, the 1.8 MeV longitude profile (or,
equivalently, the \Al \ radial profile) should be steeper than 
observed (see Fig. 5).

In summary, there is  no satisfactory explanation at present for the flux
of the 1.8 MeV line and its spatial distribution in the Milky Way.
{\it INTEGRAL} is expected to provide a more detailed spatial profile than COMPTEL
and to put more stringent limits (or, perhaps, to detect) emission from \Fh. Only when
the nature of the major \Al \ sources is clarified it will become possible 
to tackle the question of their Galactic distribution (i.e. with any yield 
dependence on metallicity - or other factors - properly taken into account).

\subsection{The \Al \ line width: a hint for mixing of SN ejecta in the ISM?}

The width of the \Al \ line was already discussed by Ramaty and Lingenfelter (1977)
who pointed outh that \Al \ ejected from SN should deccelerate in the ISM in a timescale
short compared with its decay timescale; as a consequence, the emitted \ga-ray line should
be quite narrow (narrower than the $\sim$2 keV width imposed by Galactic rotation),
making its detection relatively easy.

The HEAO-3 Ge detectors found the line to be narrow indeed: FWHM$<$3 keV 
(Mahoney et al. 1984);
however, the GRIS instrument measured a FWHM=5.4$\pm$1.4  keV, 
$\sim$3 times larger than HEAO-3
and much larger than allowed by Galactic rotation (Naya et al. 1996). 
If real, that large width can be 
interpreted either as kinematic (with the bulk of \Al \ moving
with velocities $\sim$540 km/s) or thermal (with most \Al \ atoms brough to temperatures
T$\sim$4.5 10$^8$ K). The thermal origin seems improbable, since it would imply that all
\Al \ is produced in $\sim$200 mini-starburst regions in the inner
Galaxy regions (Chen et al. 1997). A non-thermal
origin could be understood if \Al \ nuclei are incorporated in dust grains, which are
launched  by the SN explosion (Chen et al. 1997), or accelerated by the SN shock wave 
(Ellison et al. 1997) or repeatedly accelerated by SN shocks (Sturner and Naya 1999).

The SPI instrument of {\it INTEGRAL} will clarify that issue, by measuring the line width
and also the latitude distribution of the line emission. Already, {\it COMPTEL} \ measurements 
imply a vertical scaleheight of $<$220 pc for the \Al \ distribution and
suggest that the velocity of the bulk of \Al \ has not  as large a component perpendicularly
to the Galactic plane as suggested by the kinematic interpretation of the GRIS measurements
(Oberlack 1997).

\subsection{\Al \ ``hot-spots'': monitoring stars, superbubbles and young stellar associations}

The study of individual ``hot-spots'' revealed by {\it COMPTEL} \ bears on our understanding of
the evolution of young stellar associations (in the cases of Cygnus, Carina and 
Centaurus-Circinus) and even individual stars (in the case of Vela).

The Cygnus regions was studied with population synthesis models by two groups (Cervinho et al. 2001, 
Pl\"uschke et al. 2001b). The
resulting morphology of the 1.8 MeV emission compares well with the {\it COMPTEL} \ data.
However, in the case of Carina, the predicted absolute flux 
is smaller (by a factor of 5-20) than detected by
{\it COMPTEL} (Kn\"odlseder et al. 2001). 
That discrepancy may imply something interesting, either for the
(in)completeness of the stellar census of that association or for the
\Al \ yields. {\it INTEGRAL} will  establish more accurately the morphology of those
``hot-spots'' and further test the ``massive star group'' origin of \Al.

Another target of importance for future 1.8 MeV studies is the Orion/Eridanus region.
{\it COMPTEL} \ surveys of the anticenter region show significant (5 $\sigma$) extended emission
towards the south of the Orion molecular clouds. That emission could be attributed 
(Diehl 2002) to \Al \ ejected by the prominent Orion OB1 association and expanded 
into the low density cavity of the Eridanus bubble. The exansion of supernova
ejected into a previously formed cavity of peculiar shape (and not into a
medium with radial symmetry) is a novel and interesting field of study, opened by
{\it COMPTEL} \ and left for {\it INTEGRAL} to explore.

Finally, the Vela region offers the  opportunity to measure (or put upper limits
on) \Al \ yields from individual sources. The morphology of the rather extened
1.8 MeV emission detected by {\it COMPTEL} \ does not allow identification with any of the
three known objects in the field (the Vela SNR, the closest WR star \ga$^2$ Vel and 
SNR RX-J0852-4622); all three objects lie closer than 260 pc, according
to recent estimates. {\it COMPTEL} \ measurements are  compatible with current yields of SNII (in the
case of Vela SNR) and marginally compatible with current yields of \ga$^2$ Vel 
(Oberlack et al. 2000). {\it INTEGRAL} measurements in the Vela region are then
expected to place more stringent constraints on stellar models.


\section{Summary}

The aim of {\it Gamma-Ray Astronomy with Radioactivities}, as explicitly defined
by the ``founding fathers'' of the field in the 60ies (see Sec. 2) was to probe stellar
nucleosynthesis as well as supernova structure and energetics. This original aim
was reached in a spectacular way in the case of SN1987A (which, however, remains
today - and, probably, for sometime in the future -  a unique object in that respect).

On the other hand, the legacy of HEAO-3 and {\it COMPTEL} \ set new aims to the field of 
{\it Gamma-Ray Astronomy with long-lived Radioactivities}: to probe the large-scale
distribution of active nucleosynthesis sites in the Galaxy and the properties/history of any
clusterings in that distribution (young stellar associations, individual objects).
{\it INTEGRAL} is expected to perform this next step.

\section*{Acknowledgements}
 I am grateful to Roland  Diehl for his critical suggestions and comments.

\end{document}